\documentclass[aps,prb,superscriptaddress,twocolumn,showpacs,preprintnumbers,amsmath,amssymb,floatfix]{revtex4-1}
\usepackage{amssymb}
\usepackage{graphicx}
\usepackage{dcolumn}
\usepackage{bm}
\usepackage{subfigure}
\usepackage[normalem]{ulem}

\usepackage{color}
\usepackage{multirow}
\usepackage{float}
\usepackage{titlesec}
\usepackage[utf8]{inputenc}

\begin{document}

\title{Sign-reversed anomalous Nernst effect in the ferromagnetic Weyl-semimetal Fe$_{3-\delta}$GeTe$_2$: the role of Fe vacancies}

\author{Haiyang Yang}
 \thanks{These authors contributed equally.}
 \affiliation{School of Physics and Hangzhou Key Laboratory of Quantum Matters, Hangzhou Normal University, Hangzhou 311121, China}
 \affiliation{Anhui Province Key Laboratory of Condensed Matter Physics at Extreme Conditions, High Magnetic Field Laboratory, Chinese Academy of Sciences, Hefei 230031, China}

\author{Qi Wang}
 \thanks{These authors contributed equally.}
 \affiliation{School of Physics and Hangzhou Key Laboratory of Quantum Matters, Hangzhou Normal University, Hangzhou 311121, China}

\author{Junwu Huang}
 \affiliation{School of Physics and Hangzhou Key Laboratory of Quantum Matters, Hangzhou Normal University, Hangzhou 311121, China}

\author{Zhouliang Wang}
 \affiliation{School of Physics and Hangzhou Key Laboratory of Quantum Matters, Hangzhou Normal University, Hangzhou 311121, China}

\author{Keqi Xia}
 \affiliation{School of Physics and Hangzhou Key Laboratory of Quantum Matters, Hangzhou Normal University, Hangzhou 311121, China}

\author{Chao Cao}
 \email{ccao@zju.edu.cn}
 \affiliation{School of Physics and Hangzhou Key Laboratory of Quantum Matters, Hangzhou Normal University, Hangzhou 311121, China}
 \affiliation{Department of Physics, Zhejiang University, Hangzhou 310058, China}

\author{Mingliang Tian}
 \affiliation{Anhui Province Key Laboratory of Condensed Matter Physics at Extreme Conditions, High Magnetic Field Laboratory, Chinese Academy of Sciences, Hefei 230031, China}

\author{Zhuan Xu}
  \affiliation{Department of Physics, Zhejiang University, Hangzhou 310058, China}
  \affiliation{Zhejiang Province Key Laboratory of Quantum Technology and Device, Zhejiang University, Hangzhou 310027, China}

\author{Jianhui Dai}
 \email{daijh@hznu.edu.cn}
 \affiliation{School of Physics and Hangzhou Key Laboratory of Quantum Matters, Hangzhou Normal University, Hangzhou 311121, China}

\author{Yuke Li}
 \email{yklee@hznu.edu.cn}
 \affiliation{School of Physics and Hangzhou Key Laboratory of Quantum Matters, Hangzhou Normal University, Hangzhou 311121, China}


\date{\today}

\begin{abstract}
Anomalous Nernst effect, as a thermal partner of anomalous Hall effect, is particularly sensitive to the Berry curvature anomaly near the Fermi level, and has been used to probe the topological nature of quantum materials. In this work, we report the observation of both effects in the ferromagnetic Weyl-semimetal Fe$_{3-\delta}$GeTe$_2$ with tunable Fe vacancies. With decreasing Fe vacancies, the anomalous Hall conductivity evolves as a function of the longitudinal conductivity from the hopping region to the region where the intrinsic Berry curvature contribution dominates. Concomitant evolutions in the anomalous Nernst signal and the anomalous off-diagonal thermoelectric coefficient are observed below the Curie temperature, displaying a unique sign change caused by the Fe vacancies. Combining these results with first-principles calculations, we argue that the Fe-vacancy concentration plays a unique role in simultaneously tuning the chemical potential and ferromagnetism, which in turn controls the Berry curvature contribution in this family of ferromagnetic topological semimetals.

\end{abstract}


\maketitle
\section{Introduction}
In recent decades, tremendous efforts to uncover the exotic states of matter from the perspective of topology in condensed matter physics have led to the discovery of a large family of topological quantum materials such as topological insulators and topological semimetals(TSMs) \cite{PhysRevBCdAs,liang2015ultrahigh,wang2012dirac,weng2015weyl,arnold2016negative,NbP,NbAs,zhang2017electron,PRBLi,wang2019angle}. Of particular interest is the situation when magnetism meets topology in the semimetal family, where a series of unexpected transport properties, such as the anomalous Hall/Nernst effect (AHE/ANE)\cite{Mn3SnZHu,Fe3O4,ZhuPRB,CoMnGa,MPI,wangRM,CoSnSLi} and topological Hall effect (THE)\cite{PRLKanazawa} have been reported. However, whether and how these effects are related to the topological electronic structure and the broken time reversal symmetry in the material remains uncertain. Recent experimental results such as the Kagome-lattice ferromagnetic(FM) Co$_3$Sn$_2$S$_2$\cite{MPI,wangRM,CoSnSLi} and the triangular antiferromagnets Mn$_3$Sn(Ge)\cite{Mn3SnZHu,nayak2016large} have demonstrated that the ANE and AHE originate from the Berry curvature around Weyl nodes, and this connection is supported by first principles calculations(FPCs)\cite{Xu-NPJ}.

The recently discovered 2D-van der Waals magnet Fe$_3$GeTe$_2$ has been proposed as a new candidate for the FM TSM with nodal lines\cite{NM-Kim}. This material exhibits a remarkably high Curie temperature of $Tc$ = 150-220 K in the bulk samples\cite{MayFeGeTe,Zhujx-FGT,XiongYM,LiuY,Gil-FGT,YouYR}. In the monolayer, $T_c$ is even higher and approaches room temperature\cite{Yuanbozhang}; thus, this material is considered a promising new class of building blocks for heterostructures in spintronics. Furthermore, the frustrated magnetic structures due to triangular lattices formed by the iron atoms in this compound and the spin textures with spatial variation in local magnetic moments driven by the Dzlashinsky-Moriya interaction have been suggested and observed in the Fe$_3$GeTe$_2$ system\cite{XiongYM}. Such nontrivial topological spin structures are also expected to directly give rise to AHE or THE\cite{LiuY,XiongYM,YouYR}. Therefore, Fe$_3$GeTe$_2$ provides a unique platform for studying the interplay between magnetism and topology.

Compared to the Hall effect, the Nernst effect is less explored because of weak thermoelectric signals. In conventional metals, the normal contribution of the Hall effect is always finite, while the normal Nernst signal generally vanishes because of the Sondheimer cancellation\cite{Sondheimer}, indicating that the ANE may become very prominent. In fact, the thermoelectric response depends on the derivative of the conductivity and is more sensitive to the anomalous contribution of the Nernst effect. Theoretically, the AHE is dominated by the sum of Berry curvatures of all occupied bands, while the ANE is governed by the nontrivial Berry curvature at the Fermi level\cite{xiao2010berry,xiao2006berry}. Therefore, tuning the transport behavior revealed by the Nernst effect is a highly desired way to probe the topological nature of materials.

In this Letter, we report observing the AHE and ANE in bulk Fe$_{3-\delta}$GeTe$_2$($0.05 < \delta < 0.35 $), where both effects are found tunable by the Fe-vacancies. We performed transport measurements of the high-quality Fe$_{3-\delta}$GeTe$_2$ single crystals with different Fe-vacancies. We found that with decreasing Fe-vacancies the Hall conductivity $\sigma_{H}^A$ dependency on $\sigma_{xx}$ evolves from the bad-metal-hopping regime to the intrinsic regime. The anomalous Nernst signal $-S_{yx}^A$  and the anomalous off-diagonal thermoelectric conductivity $-\alpha_{yx}^A$ exhibit a prominent sign changing behavior in the low Fe-vacancy samples. Combining the FPCs, we propose that Fe-vacancies tune not only the magnetism, but also the chemical potential which in turn leads to the variation in $-\alpha_{yx}^A$ contributed from the Berry curvature anomaly when the Weyl points(or node lines) are near the Fermi level.


\section{\label{sec:level1} Methods}

Large single crystals of Fe$_{3-\delta}$GeTe$_2$($\delta = 0.05-0.35$) with a minimeter size
were grown through the flux method\cite{chen2013magnetic}.
The crystal X-ray diffraction patterns were obtained using a D/MaxrA diffractometer with CuK$_{\alpha}$
radiation at 300 K, which determines the crystal orientations, and their actual Fe contents (see Table 1 in SI) are determined using energy-dispersive X-ray
(EDX) spectroscopy. The (magneto)resistivity and Hall coefficient measurements were performed using the standard four-terminal method
in a commercial Quantum Design PPMS-9 system. The magnetization measurements were performed using a commercial SQUID
magnetometer. The thermoelectric effects, such as thermopower and the Nernst effect, were studied with a one-heater-two-thermometers technique in PPMS with a high-vacuum environment.

The first-principles calculations were performed using plane-wave basis density functional theory as implemented in the Vienna Abinit Simulation Package (VASP)\cite{kresse1996efficient}. The projected augmented wave (PAW) method\cite{PhysRevB.59.1758} and the Perdew, Burke, Ernzerhoff (PBE)\cite{PhysRevLett.77.3865} type exchange-correlation functional were employed. The energy cut-off was chosen to be 480 eV, while a 12$\times$12$\times$4 $\Gamma$-centered K-mesh was used to ensure convergence. All atomic coordinates and lattice constants were fully relaxed until the forces were less than 0.01 eV/\AA\ for all atoms. The magnetic moments of Fe atoms were taken to be along the $z$-direction, and spin-orbit coupling was added as a second variational method. After convergence, the two Fe atoms in the FeGe-plane has net moment of 1.55 $\mu_B$/Fe, while the other 4-four atoms had a net moment of 2.47 $\mu_B$/Fe, all aligned in the same direction.  These numbers significantly overestimate the experimentally observed saturating moments, similar to previous studies\cite{Zhujx-FGT}. This overestimation of the ordered moment is associated with lack of spin-fluctuations in DFT calculations, as shown in studies of iron-based superconductors\cite{PhysRevB.77.220506,PhysRevB.78.085104,PhysRevB.83.144512}.  Nevertheless, the DFT band structure reproduces the quasi-particle spectrum of a DFT+DMFT calculation, which is comparable to ARPES experiment\cite{NM-Kim}. The DFT electronic structure is then fitted to a 120-orbital tight-binding model including Fe-3d, Fe-4s, Ge-4s, Ge-4p, Te-5s, and Te-5p orbitals, using the maximally projected Wannier function method\cite{method:wannier90}.  The resulting Hamiltonian is then symmetrized using the WannSymm code\cite{WannSymmCode}, and the symmetrized Hamiltonian is employed to calculate $\alpha_{yx}^A$ using 72$\times$72$\times$72 K-mesh.

\section{Results and Discussion}

\begin{figure*}
\includegraphics[width=16cm]{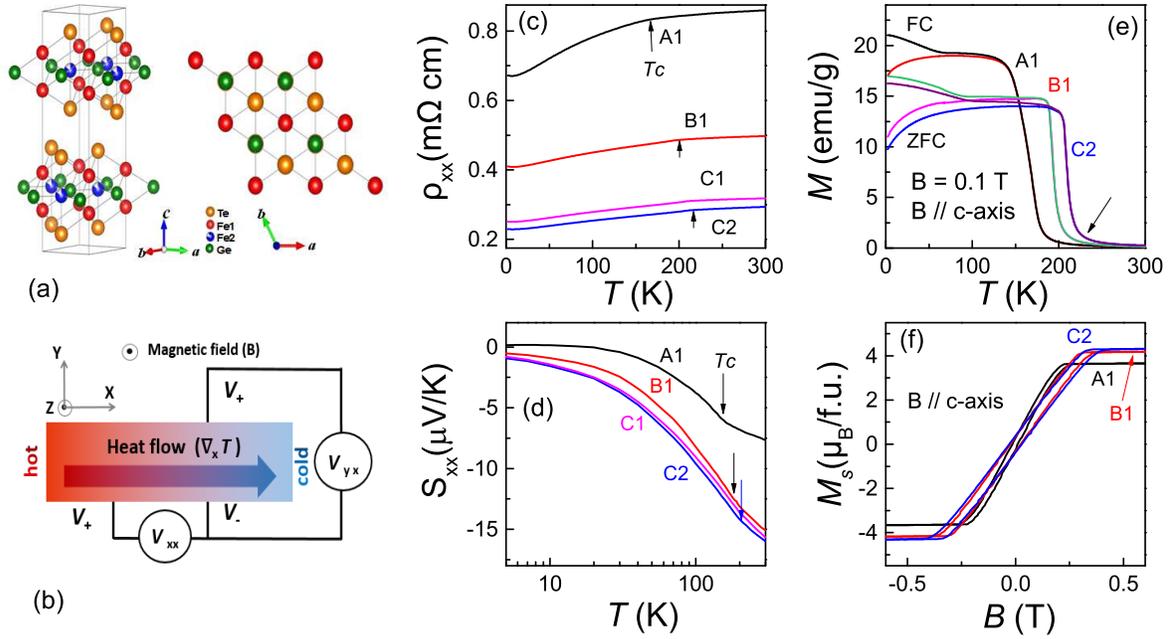}
\caption{ Crystal structure and transport property measurements of Fe$_{3-\delta}$GeTe$_2$. (a) Image of the crystal structure with the atomic positions labeled. (b) Schematic picture of the thermoelectrical measurements (Seebeck and Nernst effect). In this geometry, a temperature gradient $\nabla_x$\emph{T} produces a transverse thermoelectric voltage V$_{yx}$ in a magnetic field. Temperature dependence of the longitudinal electric resistivity $\rho_{xx}$ (c) and the longitudinal thermopower $S_{xx}$ (d). (e) Temperature dependence of magnetization with the ZFC and FC modes at B $=$ 0.1 T. (f) Saturated magnetization vs. fields at 5 K for several samples.}
\end{figure*}

A cartoon picture of the Fe$_{3-\delta}$GeTe$_2$ crystal structure (space group $P63/mmc$) with the atomic positions labeled is shown in Fig.1a, where Fe(1)-Fe(1) bonds pierce the center of each hexagon formed by Fe(2)-Ge. Using a standard two-thermometers-one-heater setup (Fig. 1b), we measured the diagonal ($\rho_{xx}$ and $S_{xx}$ ) and off-diagonal ($\rho_{yx}$ and $S_{yx}$) transport coefficients of the sample. As seen in Fig. 1c, the in-plane resistivity $\rho_{xx}$ exhibits a clear kink from 165 K to 210 K for each measured sample. The kink feature is also observed in the negative thermopower $S_{xx}(T)$ near the same temperature, as indicated in Fig.1d. As the Fe-vacancy decreases, $\rho_{xx}(T)$ decreases significantly, and $\mid S_{xx}(T)\mid$ increases monotonically, while the corresponding kink position associated with the ferromagnetic phase transition at $T_C$ shifts to high temperature.

Magnetization measurements confirmed the paramagnetic to FM phase transition at $T_C$ in Fig.1e. The temperature dependence of magnetization under the field B $\parallel$ c-axis suddenly increases below $T_C$, followed by a significant splitting at the zero field cooling (ZFC) and the field cooling (FC) curves at lower temperatures for the three samples, implying the entrance into ferromagnetic state. The lower the Fe-vacancies are, the larger the splitting of the ZFC and FC curves. This feature is not evident in the previous polycrystalline samples\cite{MayFeGeTe}. The saturated magnetization along the $c$-axis at $T =$ 5 K is displayed in Fig. 1f. The lower Fe-vacancy samples possess the larger magnetic moments (see Table 1 in SI). For instance, Sample C2 ( with the highest $T_C$ = 210 K) exhibits the largest hysteresis loop with the largest saturated magnetic moment of $M_s$ = 1.44 $\mu_B/Fe$.


\begin{figure*}
\includegraphics[width=16cm]{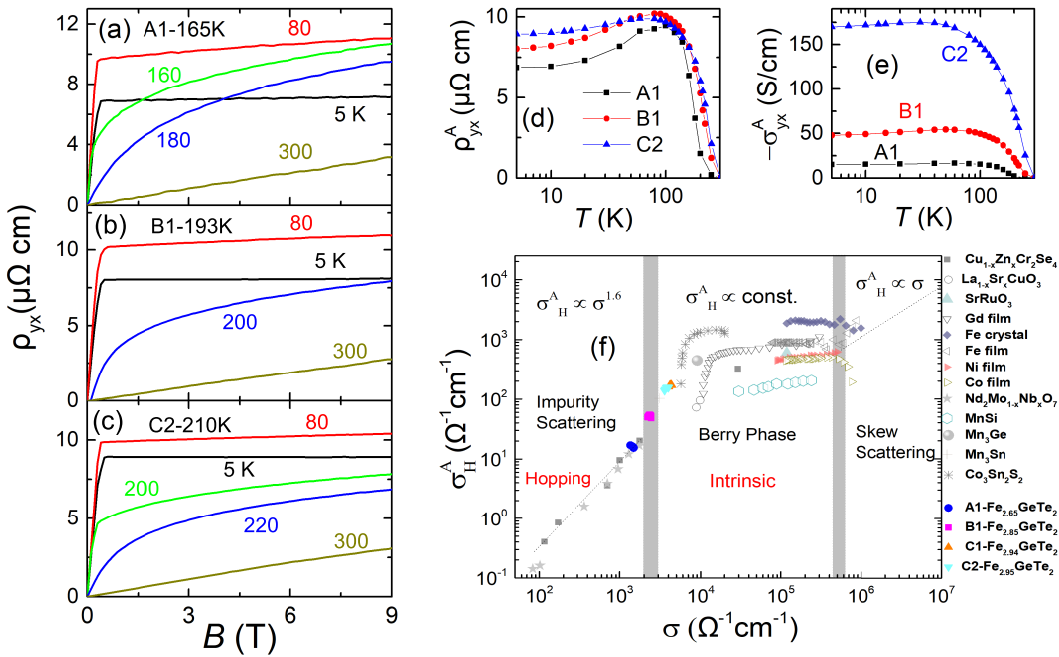}
\caption{ Hall effect of Fe$_{3-\delta}$GeTe$_2$ as B $\parallel$ c-axis $\perp$ I. Hall resistivity $\rho_{yx}$ vs. magnetic fields for Samples A1 (a), B1 (b), and C2 (c). (d) Anomalous Hall resistivity as a function of temperature; (e) temperature dependence of AHC $\sigma_{xy}$. (f) AHC $|\sigma_{H}|$ as a function of longitudinal conductivity $\sigma$ for ferromagnets, FM-TWSM Co$_3$Sn$_2$S$_2$, and Fe$_{3-\delta}$GeTe$_2$. The data of other materials were taken from references\cite{CoSnSLi}.}
\end{figure*}

Figs. 2a-2c display $\rho_{yx}$ vs. $B \parallel c \perp I$ at several temperatures below $T_C$ for Samples A1, B1 and C2, respectively. At $T$ = 5 K, the Hall resistivity $\rho_{yx}(B)$ exhibits a steep jump at low fields and then becomes almost flat at high fields, which is inconsistent with the literature\cite{LiuY}. As the temperature increases, $T_C$ ( = 165 K for A1, 193 K for B1 and 210 K for C2), $\rho_{yx}(B)$ starts to broaden and the jump is less prominent. Note that the nonlinear field dependence of $\rho_{yx}(B)$ is clearly observed above $T_C$ as well ( for more details, see SI). This AHE has been reported in some recently discovered magnetic TSMs\cite{MPI,CoSnSLi,CoMnGa}. The anomalous $\rho_{yx}^A$ is then extracted by extrapolating the high-field part of the Hall resistivity to the zero field. As shown in Fig. 2d, $\rho_{yx}^A$ increases significantly with decreasing temperature (below $T_C$), then reaches a local maximum near 80-100 K (defined as $T_{max}$) and saturates gradually at the low temperature limit.
Accordingly, the anomalous Hall conductivity (AHC) $-\sigma_{yx}^A$ can be extracted from $\rho_{yx}^A$ in Fig. 2e, with the former being calculated using the formula $-\sigma_{yx}^A= \rho_{yx}^A/({\rho_{yx}^A}^2+\rho^2)$.  $-\sigma_{yx}^A(T)$ remains almost constant below $T_{max}$. Among them, the $-\sigma_{yx}^A$ of sample C2 attains its maximum value of 175 $\Omega^{-1} cm^{-1}$ below $T_C$, close to the intrinsic contribution $-\sigma_{yx}\approx \frac{e^2}{ha_z}=235 \Omega^{-1} cm^{-1}$ ($a_z$ is the lattice constant along the $c$-axis). This value is almost one or two orders in magnitude larger than that of sample A1. In contrast, $\rho_{yx}^A$ of those samples is comparable in the whole temperature regime. Such sample-dependent diversity of the AHC is unexpected, indicating the significant role played by Fe vacancies in tuning the AHE in Fe$_{3-\delta}$GeTe$_2$.

On the basis of the unified theoretical picture\cite{PRLPureFe,PRL97}, the origin of the AHE is ascribed to an extrinsic or intrinsic mechanism due to magnetic impurity scattering or
nontrivial Berry curvature, respectively. The former is composed of the side-jump and the skew scattering. The intrinsic AHE can be expressed as $\sigma_{xy}^A = -\frac{e^2}{\hbar}\int\frac{d\mathbf{k}^3}{2\pi^3}\Omega_{z}(k)f(k)$, where $f(k)$ is the Fermi-Dirac distribution function and $\Omega_{z}(k)$ denotes the Berry curvature. These contributions can be identified by plotting $\sigma^A_H$ as function of the longitudinal conductivity $\sigma$, as shown in Fig.2f for the measured samples. Samples C1 and C2 completely fall into the intrinsic regime where $\sigma^A_H$ is dominated by the Berry-phase curvature while Sample A1 lies in the extrinsic regime where $\sigma^A_H$ is determined by the impurities scattering, consistent with the smaller $-\sigma_{yx}^A$\cite{PRL97}. As for Sample B1, it is located at the boundary between the intrinsic and extrinsic regimes. These results reveal an interesting Fe-vacancy-driven crossover from extrinsic to intrinsic mechanisms of the AHE due to the interplay of magnetism and topology. Therefore, with increasing Fe-vacancies, the decreasing moments will give rise to a decreasing intrinsic Hall effect that is proportional to magnetization. The disorder may play a minor role in contributing to the extrinsic AHE. However, the shift in chemical potential near the Fermi level may be less important for the intrinsic AHE, which is governed by the sum of the Berry curvatures of all occupied bands\cite{CoSnSLi}.

\begin{figure}
\includegraphics[angle=0,width=8cm,clip]{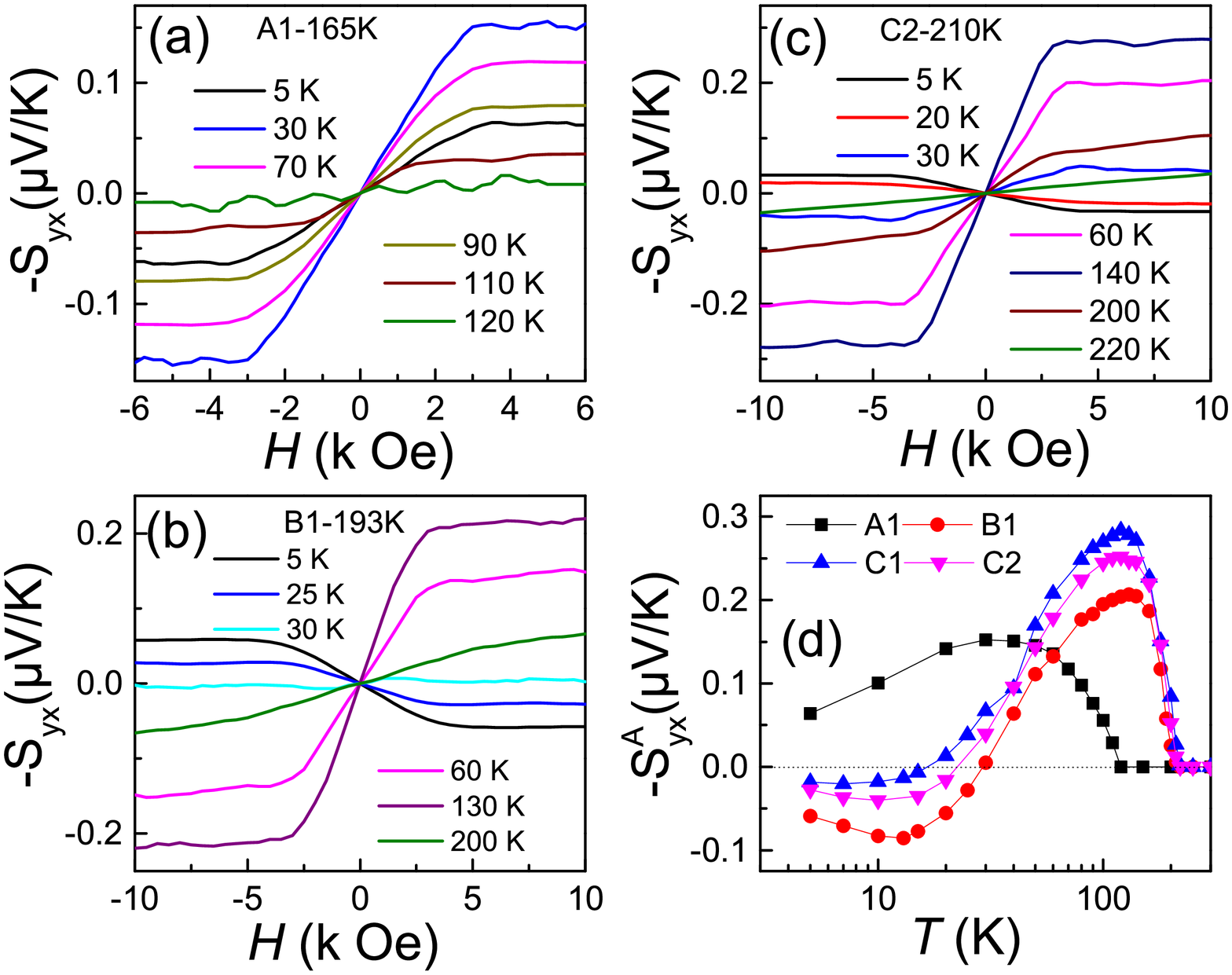}
\caption{Anomalous Nernst effect in Fe$_{3-\delta}$GeTe$_2$. Field dependence of the Nernst signal -S$_{yx}$ for Samples A1(a), B1 (b) and C1 (c). (d) Temperature-dependence of the ANS -S$_{yx}^A$ for the measured samples.}
\end{figure}

Compared with the AHE, our samples exhibit a more exotic ANE in Fig. 3a-3c. Below $T_C$, the Nernst signal (NS), $-S_{yx}$, shows a clear jump at low fields and tends to saturate at high fields. Above $T_C$, the abnormal jump is less identified, and $-S_{yx}$ obeys a linear field dependence, indicating a normal NS. Surprisingly, $-S_{yx}$ in Samples B1 and C2 follows a positive field dependence($\propto B$) above 20-30 K but reverses to a negative one ($\propto - B$) below this temperature, revealing a remarkable sign change in $-S_{yx}$ with decreasing temperature. This behavior is absent in Sample A1. Such unusual field dependence in a Nernst signal does not match the magnetization curve and the Hall resistivity data in a simple way. Rather, it indicates a strong dependence of the ANE on the amount of Fe vacancies. The anomalous NS $-S_{yx}^A(T)$ is then extracted from the $-S_{yx}(B)$ data (for detailed information, see SI) as displayed in Figure 3d. $-S_{yx}^A(T)$ clearly increases suddenly below $T_C$, and then shows a broad peak near 30 K for Sample A1, 130 K for Sample B1, and 140 K for Samples C1 and C2. The maximum value of $-S_{yx}^A$ varies from 0.15 to 0.3 $\mu$ V/K, which is much larger than those of conventional magnets (the pure metal Fe\cite{PRLPureFe}, CuCr$_2$Se$_{4-x}$Br$_x$\cite{PRLCuCrSe}, and the single crystal Fe$_3$O$_4$\cite{Fe3O4}), and is comparable to the magnetic TSM (the Mn$_3$Sn(Ge)\cite{Mn3SnZHu,nayak2016large}). The unexpected crossover from positive to negative $-S_{yx}^A$ in Samples B1, C1 and C2 occurs near 15-25 K, followed by a local minimum at lower temperatures of $\sim$ 10-15 K. To date, this sign reversing feature in ANE has not been reported in the known magnetic TSMs.


\begin{figure}
\includegraphics[angle=0,width=8cm,clip]{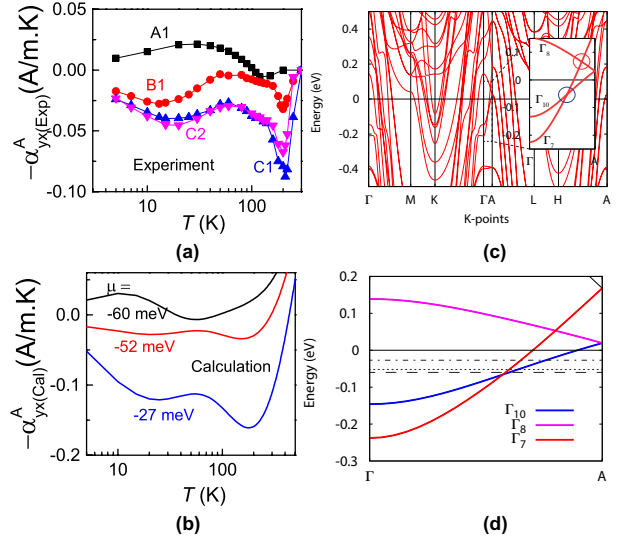}
\caption{Band structure and $-\alpha^A_{yx}$ in experiment and calculation. (a) $-\alpha^A_{yx}$ as a function of temperature according to our experimental data. (b) Calculated $-\alpha_{yx}^A$ vs. temperature.(c) Energy band of sample with SOC. (d) Enlarged band structure with the shift in chemical potential.}
\end{figure}

We now turn to the anomalous off-diagonal thermoelectric conductivity $\alpha_{yx}^A(T)$, which is closely associated with $S^A_{yx}$ and $\sigma^A_{yx}$ via $\sigma_{xx}$ and $S_{xx}$ by
$-\alpha^A_{yx} = \sigma^A_{xy}S_{xx}+S^A_{xy}\sigma_{xx}$ (for more details, see SI).
As shown in Fig. 4a, $\mid\alpha_{yx}\mid$ is moderate in the range of 0.01-0.1 Am$^{-1}$K$^{-2}$. It displays a nonmonotonic temperature-dependence below $T_C$. For Sample A1, $-\alpha^A_{yx}(T)$ is positive with a broad hump near 30 K, then approaches zero at zero temperature. In contrast, $-\alpha_{yx}^A(T)$ in Samples B1, C1, and C2 becomes negative and exhibits a local minimum near 20 K and a dip near 200 K. The dip position is slightly above the Curie temperature, and linked to the ignorable Nernst signal because of the Sondheimer cancellation in the paramagnetic state of a normal metal\cite{Sondheimer}. The local minimum is due to $-S_{yx}^A$ vanishing when the sign-reversal occurs such that the anomalous Hall conductivity is the main contributor to $-\alpha^A_{yx}$ ($=\sigma^A_{xy}S_{xx}$).

To clarify this point, we performed an FPC for the pristine compound and obtained $-\alpha_{yx}^A$ using semiclassical approach\cite{xiao2010berry}(for details, see SI), as shown in Fig. 4b. The calculated band structure with spin-orbit coupling (SOC) is shown in Fig.4c, where the band inversions from the $\Gamma$-point to the $A$-point near the Fermi level yield two crossing points (W1 and W2) located at -65 meV and 70 meV, respectively. Notably that the chirality of W1 is $\pm2$, while W2 is the normal Weyl point with chirality of $\pm1$. Moreover, the chiralities of the two nodal points sitting on the same side of the $k_x$-$k_y$ plane have different signs. Since $\alpha^A_{yx}=-\frac{e}{T\hbar}\int \frac{d\mathbf{k}^3}{2\pi^3}\Omega_{n,z}(k)s_{nk} $, where $s_{nk}=(\epsilon_{n,k}-\mu)f_{n,k}+k_BT \ln[1+e^{-\beta(\epsilon_{n,k}-\mu)}]$, the ANE $\alpha^A_{yx}$ is related to the Berry curvature anomalies within thermal thinness. As a result, the Weyl nodes with opposite signs compete with each other when the chemical potential is tuned in between. Govern that the reduction of the saturated moment is roughly proportional to the vacancy concentration $x$, and that previous coherent potential approximation calculations suggest a negligible change in spin-splitting size in systems with iron vacancies\cite{NM-Kim}, we argue that the major effect of iron-vacancy in our cases (up to $\sim$10\%) is charge doping without altering the local moment on individual Fe atoms. Therefore, we simulate the influence of Fe-vacancies by rigidly shifting the chemical potential, i.e. ignoring the local moment change and focusing on the charge doping effect.
Three representative chemical potentials (-60 meV, -52 meV, and -27 meV) are chosen as indicated in Fig. 4d and the corresponding value of $-\alpha_{yx}^A$ is calculated as shown in Fig. 4b.
The calculated $-\alpha_{yx}^A$ does evolve from positive to negative below $T_C$ under the shift in the chemical potential from -60meV to -27meV. All calculated results (in Fig. 4b) and experimental results (in Fig. 4a) are in fair agreement. The theoretical values of $-\alpha_{yx}^A$ are slightly larger than the
experimental values possibly because of the neglected influence of the temperature-dependent magnetic moments as well as the contribution from other possible mechanisms.
The good agreement between calculation and experiment confirms that the dominant contribution to $-\alpha_{yx}^A$ is from the Berry curvature near the Weyl points close to the Fermi level. The competition between two Weyl nodes with different topological charges caused by the doping effect gives rise to the nontrivial behavior of the ANE.


In short, the AHE and ANE exhibit a clear difference in transport with respect to the Fe vacancy concentration in the system. At high Fe vacancy levels as in Sample A1, the system exhibits low $T_C$ and small $\mu_{sat}$, associated with small $-\sigma_{yx}^A$ and $-\alpha_{yx}^A$ caused by the intrinsic magnetic structure or the extrinsic disorder effect. At low Fe vacancy levels as in Sample C2, the system displays high $T_C$ and large $\mu_{sat}$, associated with large $-\sigma_{yx}^A$ and $-\alpha_{yx}^A$. $-S_{yx}^A(T)$ undergoes a sign change in the low temperature regime where $-\alpha_{yx}^A(T)$ has the opposite sign below $T_C$. The sign changing behavior can be attributed to the shift in the chemical potential which can tune the contribution of Berry curvature in the electronic band structure.

\section{Conclusion}

In summary, we studied the AHE and the ANE of bulk Fe$_{3-\delta}$GeTe$_2$($0.05 < \delta < 0.3 $) compounds. We found that with decreasing Fe-vacancy, $-\sigma_{yx}^A$ dependency on $\sigma_{xx}$ evolves from the bad metal hopping regime to the intrinsic regime, indicating the different dominating mechanisms behind the AHE. The accompanying $-S_{yx}^A(T)$ undergoes an unexpected sign change at low temperature, where $-\alpha_{yx}^A(T)$ has the opposite sign below $T_C$. Calculation and experimentation suggest that the Fe-vacancies can not only tune the magnetism, but also tune the chemical potential, leading to an interesting variation ins $-\alpha_{yx}^A(T)$ whose dominating contribution is from the Berry curvature when the Weyl points (or nodal lines) are near the Fermi level. The results also indicate a possible crossover from extrinsic to intrinsic AHE and ANE driven by the Fe-vacancies. Our work reveals the unique role played by Fe-vacancies in the present ferromagnetic TSMs and suggests that this family could be a new platform for studying the interplay between magnetism and topology.

\section*{Acknowledgments}
The authors thank Yongkang Luo and Huiqiu Yuan for fruitful discussions and valuable comments. This research was supported in part by the NSF of China (under the Grant Nos. U1932155, 11874136, 11874137 and U19A2093), the National Key Projects for Research and Development of China (Contract No. 2019YFA0308602), and the Key R\&D Program of Zhejiang Province China (No. 2021C01002). Yu-Ke Li was also supported by an open program from Wuhan National High Magnetic Field Center (2016KF03).


%

\end{document}